\documentclass[twocolumn]{aastex631}
\usepackage{txfonts}

\submitjournal{ApJL}

\begin{document}

\title{Evidence for Nightside Water Emission Found in Transit of Ultrahot Jupiter WASP-33b}

\correspondingauthor{Guo Chen}
\email{guochen@pmo.ac.cn}

\author[0009-0001-5682-5015]{Yuanheng Yang}
\affiliation{CAS Key Laboratory of Planetary Sciences, Purple Mountain Observatory, Chinese Academy of Sciences, Nanjing 210023, China}
\affiliation{School of Astronomy and Space Science, University of Science and Technology of China, Hefei 230026, China}

\author[0000-0003-0740-5433]{Guo Chen}
\affiliation{CAS Key Laboratory of Planetary Sciences, Purple Mountain Observatory, Chinese Academy of Sciences, Nanjing 210023, China}
\affiliation{CAS Center for Excellence in Comparative Planetology, Hefei 230026, China}

\author[0000-0001-9585-9034]{Fei Yan}
\affiliation{Department of Astronomy, University of Science and Technology of China, Hefei 230026, China}

\author[0000-0003-2278-6932]{Xianyu Tan}
\affiliation{Tsung-Dao Lee Institute \& School of Physics and Astronomy, Shanghai Jiao Tong University, Shanghai 201210, China}

\author[0000-0002-9260-1537]{Jianghui Ji}
\affiliation{CAS Key Laboratory of Planetary Sciences, Purple Mountain Observatory, Chinese Academy of Sciences, Nanjing 210023, China}
\affiliation{CAS Center for Excellence in Comparative Planetology, Hefei 230026, China}



\begin{abstract}

To date, the dayside thermal structure of ultrahot Jupiters (UHJs) is generally considered to be inverted, but their nightside thermal structure has been less explored. Here we explore the impact of nightside thermal emission on high-resolution infrared transmission spectroscopy, which should not be neglected, especially for UHJs. We present a general equation for the high-resolution transmission spectrum that includes planetary nightside thermal emission. This provides a new way to infer the thermal structure of the planetary nightside with high-resolution transmission spectroscopy. Using the cross-correlation technique, we find evidence for the presence of an H$_2$O emission signature on the UHJ WASP-33b during the transit, indicating an inverted temperature structure on its nightside. Such a result suggests a stronger heat transport through the circulation than currently expected. An alternative explanation is that the rotating visible hemisphere during transit leads to the potential contribution of the limb and dayside atmospheres to the detected emission signature. In the future, the combination of high-resolution full-phase curve spectroscopic observations and general circulation models will hopefully solve this puzzle and provide a complete picture of the three-dimensional nature of the chemistry, circulation, and thermal structure of UHJs.

\end{abstract}


\keywords{\href{http://astrothesaurus.org/uat/487}{Exoplanet atmospheres (487)}; 
\href{http://astrothesaurus.org/uat/2021}{Exoplanet atmospheric composition (2021)}; 
\href{http://astrothesaurus.org/uat/2133}{Transmission spectroscopy (2133)};
\href{http://astrothesaurus.org/uat/2096}{High resolution spectroscopy (2096)}; 
\href{http://astrothesaurus.org/uat/753}{Hot Jupiters (753)}; 
\href{http://astrothesaurus.org/uat/498}{Exoplanets (498)};
\href{http://astrothesaurus.org/uat/509}{Extrasolar gaseous giant planets (509)}
}


\section{Introduction}
	
Ultrahot Jupiters (UHJs) are gas giant planets that orbit very close to their host stars with high dayside temperatures \citep[$T_\mathrm{day}$ $\ge$ 2200\,K,][]{2018A&A...617A.110P}. The intense stellar irradiation on the dayside of UHJs can lead to thermal dissociation of key radiatively active molecules, which are then transported to the nightside for recombination. This process results not only in global chemical inhomogeneities but also in significant changes in the radiative balance, leading to remarkable alterations in the dynamical pattern and thermal structure. Due to their observational accessibility, UHJs serve as an ideal laboratory for testing planetary atmospheric physics and models under extreme conditions.
	
General circulation models (GCMs) play a crucial role in deciphering the chemical and physical processes within the three-dimensional (3D) atmospheres of UHJs. In particular, they provide insights into the dynamics and temperature-pressure (T-P) structures. The day-night temperature differences in hot Jupiters could be controlled by various factors, including stellar irradiation, circulation patterns, radiative heating/cooling, and frictional drag \citep{2002A&A...385..166S,2016ApJ...821...16K,2019ApJ...886...26T}. In addition to the day-night temperature differences, the vertical temperature structures also vary from the dayside to the nightside. Typically, the temperature decreases with decreasing pressure (i.e., non-inverted T-P). However, several studies \citep{2003ApJ...594.1011H,2005ApJ...627L..69F,2008ApJ...678.1419F} proposed that the T-P of planetary atmospheres strongly irradiated by stars bifurcates, especially in the presence of optical absorbers such as TiO and VO, resulting in a thermal inversion, i.e., the temperature increases with decreasing pressure. \citet{2015A&A...574A..35P} further presented that the non-gray thermal effects, which reduce the cooling capacity of the upper atmosphere, are the reason for the thermal inversion. Furthermore, metal atoms, molecules (such as silicon oxide and metal hydrides), and H$^-$ can also induce dayside temperature inversions in extremely irradiated hot Jupiters \citep{2018ApJ...866...27L,2018ApJ...855L..30A,2018A&A...617A.110P}. \citet{2019MNRAS.485.5817G} showed that some refractory species other than TiO and VO or low infrared opacity can also cause thermal inversion. While current theories satisfactorily explain the dayside temperature inversion, there remains a lack of theory and observational constraints on the vertical temperature structure of the nightside atmosphere, which is empirically assumed to be either non-inverted or isothermal.

In recent years, transmission spectroscopy has revealed an increasing number of chemical species in exoplanet atmospheres, ranging from neutral or weakly ionized atoms to molecules. It typically assumes a dark planet transiting the host star and considers only the chromatic absorption of stellar light by planetary atmospheres. Several studies \citep{2010MNRAS.407.2589K,2020ApJ...898...89C,2020AJ....160..197M,2021AJ....161..174M} have attempted to investigate the effect of the nightside emission from hot Jupiters on the transit light curves and found that the chromatic dilution of the mid-infrared transit depth can be as large as 10$^{-4}$, which may bias the inference of chemical abundances and thermal structures of some exoplanet atmospheres in the era of the James Webb Space Telescope. However, the effect of velocity-resolved nightside emission on transmission spectroscopy remains unexplored. 

In this letter, we focus on the implications for high-resolution infrared transmission spectroscopy of wavelength-dependent thermal emission arising from diverse temperature structures on the nightside of UHJs. We take WASP-33b as an example for illustration, which is a UHJ with a mass of 2.2~$M_\mathrm{J}$ and a radius of 1.5~$R_\mathrm{J}$ orbiting a $\delta$-Scuti A-type star with a period of 1.22 days \citep{2010MNRAS.407..507C}.
Despite being one of the most intensely irradiated hot Jupiters, WASP-33b exhibits heat transport efficiency comparable to other hot Jupiters \citep{2018AJ....155...83Z}. Instead of limiting ourselves to the nightside thermal structures predicted by the current mainstream GCMs, we infer possible nightside thermal structures directly from the observational data and try to understand them in order to contribute to the theoretical development of 3D circulation patterns and energy transport efficiencies in UHJs. 

\section{Radiation Model of the Nightside Atmosphere of Planets}\label{RM}
	
\subsection{Planet's Thermal Emission Signal vs. Transmission Absorption Signal}
	
The planetary atmosphere can absorb light from the host star during transit. Each atomic or molecular species has a different extinction coefficient for the stellar spectrum at a given wavelength. Therefore, the information about the species of the planetary atmosphere is imprinted on the outgoing stellar spectrum during the transit. The classical transmission spectrum ($T_\mathrm{m}$) is defined as the spectral flux ratio between out-of-transit and in-transit, given by (omitting the notation $\mathrm{\lambda}$), 
\begin{eqnarray}
    T_\mathrm{m} = 1-\frac{F_\mathrm{out}-F_\mathrm{in}}{F_\mathrm{out}} = 1-\frac{R_\mathrm{eff}^2 \cdot \psi_\mathrm{m} (u)}{R_\mathrm{s}^2}, \nonumber
\end{eqnarray}
where $R_\mathrm{eff}$ is the effective planetary radius as a function of wavelength, $R_\mathrm{s}$ is the stellar radius, and $\psi_\mathrm{m}$ is the planetary rotational broadening profile for the transmission spectrum. We computed the rotational broadening kernel of the transmission spectrum according to the method of \citet{2023MNRAS.522.5062B} and modified the code developed by \citet{2023RNAAS...7...91C} to implement fast rotational broadening over a wide range of wavelengths. We initially set the thickness of the atmosphere to $z=5H$, where $H$ is the atmospheric scale height and tidal locking is assumed.

In addition to absorbing starlight from their host stars, planetary atmospheres also produce their own thermal emission. As the thermal emission travels from the inner to the outer atmosphere, it interacts with local matter. The spectral line features in the outgoing spectrum are determined by the vertical T-P profile and the chemical abundance profiles of the planetary atmosphere. The thermal emission spectrum ($T_\mathrm{e}$) can be expressed as (omitting the notation $\mathrm{\lambda}$), 
\begin{eqnarray}
    T_\mathrm{e} = \frac{F_\mathrm{p} \cdot R_\mathrm{p}^2 \cdot \psi_\mathrm{e}(u)}{F_\mathrm{s} \cdot R_\mathrm{s}^2}, \nonumber
\end{eqnarray}
where $F_\mathrm{p}$ is the thermal emission flux at the top of planetary atmosphere, $F_\mathrm{s}$ is the stellar flux, $R_\mathrm{p}$ is the planetary radius, $R_\mathrm{s}$ is the stellar radius, and $\psi_\mathrm{e}$ is the rotational broadening profile for the thermal emission. Unlike the transmission spectrum, here we use the code from \citet{2023RNAAS...7...91C} directly to implement fast rotational broadening.

For high-resolution observations, we focus on the strength and profile of the spectral lines, which are closely related to local temperatures and temperature gradients in the planetary atmosphere. For planets with possible nightside temperature gradients, the spectral line features of the transmission spectrum are coupled with the spectral line features of the planet's own thermal emission during transit. For WASP-33b, the magnitude of the line intensity of the $T_\mathrm{m}$ is about $10^{-4}$. If there is a sufficient temperature gradient, the magnitude of the line intensity of the $T_\mathrm{e}$ is also about $10^{-4}$. Therefore, the signals from high-resolution transmission lines and thermal emission lines may be comparable.
   
\subsection{General High-Resolution Transmission Spectrum with Planetary Nighside Emission}\label{CRMofNPA}
\begin{figure}
\centering
\includegraphics[width=\hsize]{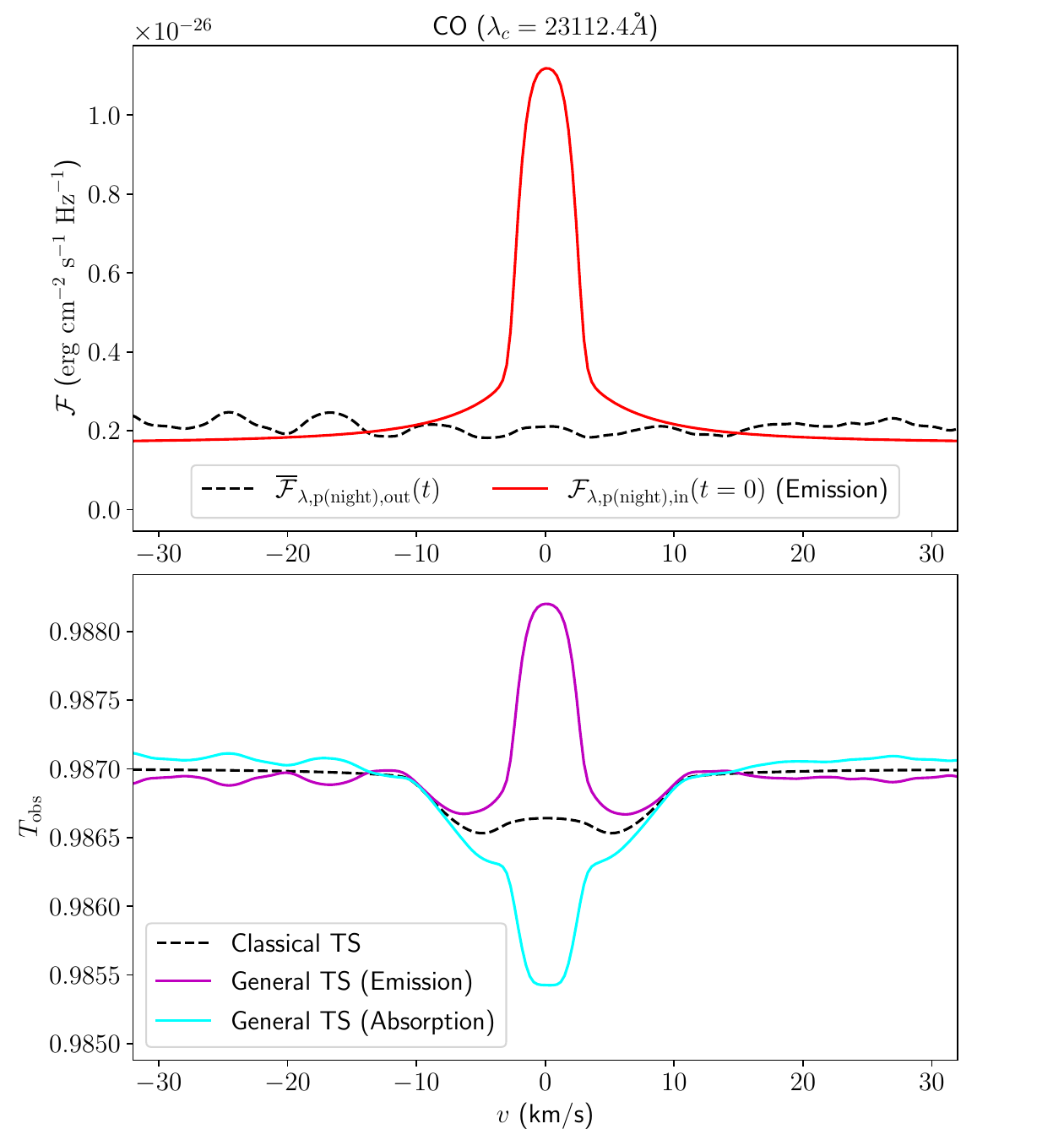}
\caption{\emph{Top panel}: Comparison between time-averaged nightside emission flux $\overline{\mathcal{F}}_\mathrm{\lambda, p(night), out}(t)$ and nightside emission flux at the moment of mid-transit $\mathcal{F}_\mathrm{\lambda, p(night), in}(t=0)$. \emph{Bottom panel}: Comparison between classical and general transmission spectrum. The gray dashed line shows the classical transmission spectrum. The solid lines show our proposed general transmission spectra, which incorporate the nightside emission. The purple and cyan colors indicate that the nightside temperature structure is inverted and non-inverted, respectively. All models include rotational broadening.}
\label{Fig0}
\end{figure}

A transmission spectrum is defined as the spectral flux difference between out-of-transit and in-transit:
\begin{equation}\label{M1}
    \Delta_{\lambda}(t) \equiv \frac{\overline{\mathcal{F}}_\mathrm{\lambda, out} - \mathcal{F}_\mathrm{\lambda, in}(t)}{\overline{\mathcal{F}}_\mathrm{\lambda, out}},
\end{equation}
where $t$ denotes time (or phase), indicating that high-resolution transmission spectra vary at different times due to Doppler shift and geometric effects (e.g., ingress/egress vs. full transit). To unpack each terms, we have 
\begin{eqnarray}\label{M2}
    &&\overline{\mathcal{F}}_\mathrm{\lambda, out} \equiv \mathcal{F}_\mathrm{\lambda, s} + \overline{\mathcal{F}}_\mathrm{\lambda, p(night), out}(t), \nonumber \\
    &&\mathcal{F}_\mathrm{\lambda, in}(t) \equiv \mathcal{F}_\mathrm{\lambda, s} - \mathcal{F}_\mathrm{\lambda, s, obscured}(t) + \mathcal{F}_\mathrm{\lambda, p(night), in}(t),
\end{eqnarray}
where ``p(night)'' denotes the planetary nightside, ``s'' denotes the star, ``out'' denotes out-of-transit, ``in'' denotes in-transit, and ``obscured'' denotes the stellar light obscured by planet (including atmosphere and disk). Substituting Equation~(\ref{M2}) into Equation~(\ref{M1}), we have
\begin{equation}\label{A4}
    \Delta_\mathrm{\lambda}(t) = \frac{\mathcal{F}_\mathrm{\lambda, s, obscured}(t) + \overline{\mathcal{F}}_\mathrm{\lambda, p(night), out}(t) - \mathcal{F}_\mathrm{\lambda, p(night), in}(t)}{\mathcal{F}_\mathrm{\lambda, s} + \overline{\mathcal{F}}_\mathrm{\lambda, p(night), out}(t)}.
\end{equation}
For high-resolution transmission spectroscopy, taking into account the Doppler effect from the variations in the line-of-sight velocity (see top panel of Figure~\ref{Fig0}), $\overline{\mathcal{F}}_\mathrm{\lambda, p(night), out}(t) \neq \mathcal{F}_\mathrm{\lambda, p(night), in}(t)$.

\begin{figure*}
\centering
\includegraphics[width=\hsize]{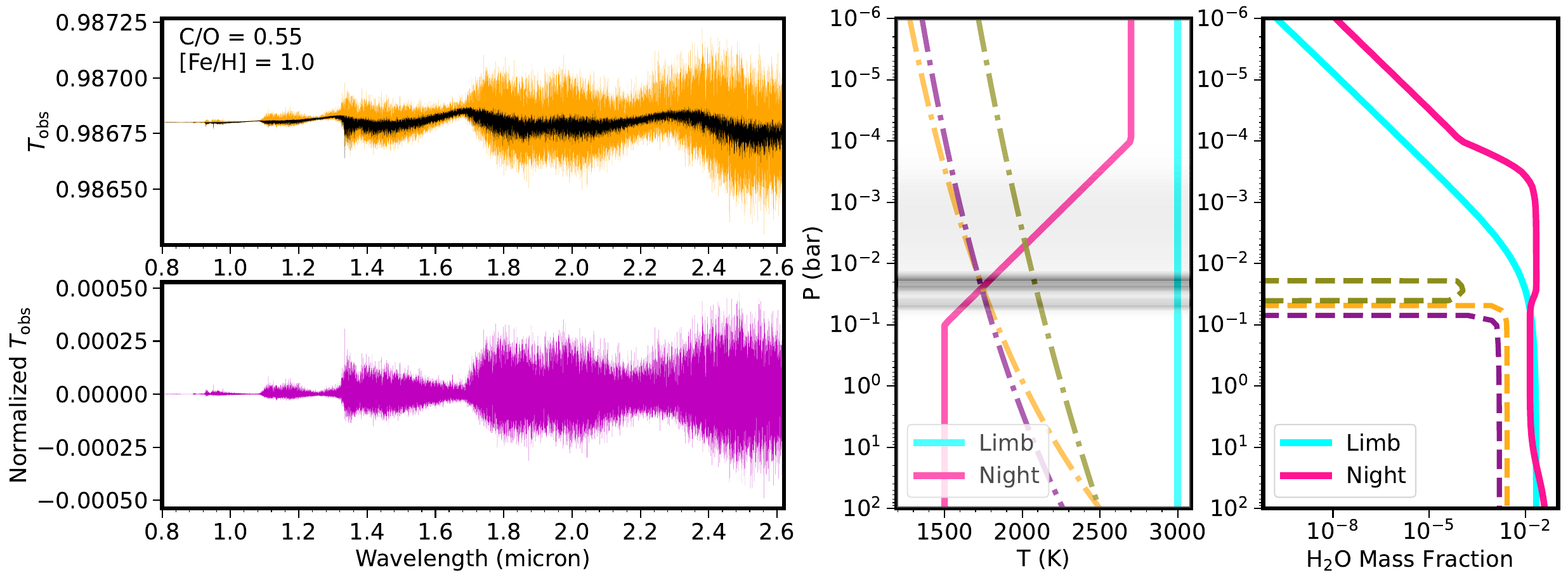}
\caption{The H$_2$O transmission spectrum model for WASP-33b. The upper left panel shows the model spectra of H$_2$O, our general transmission spectrum (with nightside emission) in orange and the classical transmission spectrum (without nightside emission) in black. The lower left panel shows the normalized continuum-free model spectra. The middle panel shows the T-P profiles for the limb and nightside atmospheres. The gray shaded area shows the contribution function of the nightside thermal emission, integrated over the wavelength space. The dotted dashed lines show the condensation profiles of condensates commonly expected on high-temperature giants \citep[][]{2010ApJ...716.1060V,2017MNRAS.464.4247W}. The peak of the contribution function here is above the intersection of the condensation profile with the nightside T-P, indicating that the thermal emission is not obscured by clouds. The right panel shows the H$_2$O mass fraction calculated from the chemical equilibrium given by the T-P in the middle panel, compared to that of the nightside condensates. 
}
\label{Fig1}
\end{figure*}	

We show in Appendix~\ref{DGE} the derivation of the general equation for high-resolution transmission spectrum with nighside emission. The one-dimensional (1D) general transmission spectrum (or called ``transit depth'') can be expressed as,
\begin{eqnarray}\label{MainEqu}
    \Delta_{\lambda}(t)=\left[\varepsilon_\mathrm{\lambda, het} \frac{A_\mathrm{\lambda, eff}(t)}{A_\mathrm{s}}-\frac{L_\mathrm{\lambda, p(night)}(t)}{\bar{I}_\mathrm{\lambda, s}} \frac{A_\mathrm{\lambda, eff}(t)}{A_\mathrm{s}}\right] \cdot d_\mathrm{\lambda, night}, 
\end{eqnarray}
where 
$d_\mathrm{\lambda, night} = 1/[1+\overline{\mathcal{F}}_\mathrm{\lambda, p(night), out}(t)/\mathcal{F}_\mathrm{\lambda, s, out}] = 1/(1+F_\mathrm{p} R_\mathrm{p}^2/F_\mathrm{s} R_\mathrm{s}^2)$ is the ``nightside emission dilution factor'' \citep{2010MNRAS.407.2589K}. 
$\varepsilon_\mathrm{\lambda, het}$ is the ``unocculted stellar heterogeneities factor'' \citep{2022ApJ...929...20M}. $A_\mathrm{\lambda, eff}$ and $A_\mathrm{s}$ denote the planetary effective area and stellar area, respectively. $L_\mathrm{\lambda, p(night)} = I_\mathrm{\lambda, p(night)}-\bar{I}_\mathrm{\lambda, p(night)}$ and $\bar{I}_\mathrm{\lambda, s}$ are the approximate spectral line intensity and mean stellar intensity, respectively. 
$L_\mathrm{\lambda, p(night)}(t)$ is the line intensity of the nightside emission at a given time after subtracting the continuum spectrum. For low-resolution transmission spectrum, $L_\mathrm{\lambda, p(night)} = 0$, therefore, Equation~(\ref{MainEqu}) turns into Equation~(1) of \citet{2022ApJ...929...20M}.

In practice, assuming $\varepsilon_\mathrm{\lambda, het}=1$, we can rewrite Equation~(\ref{MainEqu}) as (omitting the notations $t$ and $\mathrm{\lambda}$),
\begin{eqnarray}
    &T_\mathrm{obs} &= 1-\Delta \nonumber \\
    &&= 1-\frac{R_\mathrm{eff}^2 \cdot \psi_\mathrm{m}(u)}{R_\mathrm{s}^2} \left\{1-\frac{\mathcal{G} \left[F_\mathrm{p} \cdot \psi_\mathrm{e}(u) \right]}{F_\mathrm{s}}\right\} \cdot d_\mathrm{\lambda, night}, 
\end{eqnarray}
where $T_\mathrm{obs}$ denotes the general transmission spectrum model for subsequent analyses, and $\mathcal{G}$ denotes Gaussian high-pass filter. $R_\mathrm{s}$ and $F_\mathrm{s}$ denote the stellar radius and flux (blackbody assumed), respectively. We use petitRADTRANS \citep{2019A&A...627A..67M} to implement the radiative transfer to compute $R_\mathrm{eff}$ and $F_\mathrm{p}$. We set different two-point temperature structures for the $R_\mathrm{eff}$ and $F_\mathrm{p}$ calculations to account for the longitudinal variation in the temperature structure of the planetary atmosphere. The abundances of the chemical components are calculated from chemical equilibrium and solar elemental abundances under the assumed temperature structure, carbon-oxygen ratio (C/O), and metallicity ([Fe/H]). Our model also includes the continuum opacity of H$^-$ as well as the effects of condensation clouds. The bottom of Figure~\ref{Fig0} presents an example that compares the general transmission spectrum with the classical transmission spectrum.
   
\section{Observations and Methods}\label{OandM}

\subsection{Observations and Data Reduction}

We collected archived data observed by CARMENES \citep{2018SPIE10702E..0WQ} and GIANO-B \citep{2017EPJP..132..364C} for the transits of WASP-33b from the Calar Alto Archive\footnote{\url{http://caha.sdc.cab.inta-csic.es/calto/jsp/searchform.jsp}} and the Italian Center for Astronomical Archive\footnote{\url{http://archives.ia2.inaf.it/tng/}}, respectively. The CARMENES spectrograph is mounted on the 3.5 m telescope at Calar Alto Observatory. The CARMENES near-infrared channel has a wavelength coverage of 0.96--1.71 $\micron$ with high spectral resolution ($R\sim 80,400$). The GIARPS \citep{2016SPIE.9908E..1AC} instrument, consisting of the GIANO-B spectrograph and the HARPS-North spectrograph, is installed on the Telescopio Nazionale Galileo, allowing simultaneous observation in the visible and near-infrared wavelengths. The GIANO-B spectrograph has a high resolution of $R\sim 50,000$ and covers the wavelength range from 0.95 to 2.45 $\micron$. The basic method of high-resolution spectral data reduction follows \citet{2019A&A...625A.107G}, \citet{2021Natur.592..205G}, \citet{2022A&A...659A...7Y} and \citet{2024AJ....167...36Y}. We used SYSREM \citep{2005MNRAS.356.1466T} to remove telluric contamination and stellar spectral lines to obtain the final order-by-order spectral matrices for subsequent cross-correlation analyses (see Appendix~\ref{Appendix_DR} for the detailed data reduction process). The algorithm was introduced to analyze high-resolution spectroscopic time series by \citet{2013MNRAS.436L..35B} and has become a standard techqniue in atmospheric characterization of (ultra-)hot Jupiters \citep[e.g.,][]{2017AJ....153..138B,2017AJ....154..221N,2018ApJ...863L..11H,2020MNRAS.493.2215G,2020A&A...640L...5Y}.
	
\subsection{Cross-Correlation}

To search for the molecular spectral signal, we implemented the cross-correlation technique \citep{2010Natur.465.1049S}. The transmission spectrum models (see Section~\ref{CRMofNPA}) were convolved with a Gaussian function to match the instrumental resolution and resampled to the same wavelength grid as the observed spectra. The models were further normalized with a 12-point Gaussian high-pass filter to remove the broadband spectral features. The models were shifted from $-$150 to $+$150 km~s$^{-1}$ with a step of 3 km~s$^{-1}$, which is larger than the pixel scale of both instruments to avoid oversampling. At each spectral shift, we computed the weighted cross-correlation function (CCF) by multiplying the weighted residual spectrum with the shifted spectral model. The CCF is formulated as:
\begin{equation}
    \mathrm{CCF}=\sum \frac{r_{i} m_{i}(\varv)}{\sigma_{i}^{2}},
\end{equation}
where $r_{i}$ represents the residual spectrum, $m_{i}(\varv)$ is the spectral model shifted by velocity $\varv$, and $\sigma_{i}$ denotes the error at the wavelength point $i$. For each residual spectrum, a corresponding CCF was derived. These individual CCFs were then combined by stacking to generate the CCF map for each observation and model.

To enhance the signal-to-noise ratio (S/N) of the planetary signals, the CCF map was converted from the terrestrial rest frame to the planetary rest frame using a grid of assumed planetary orbital velocity semi-amplitudes ($K_\mathrm{p}$). Assuming the planet has a circular orbit, the radial velocity (RV) of the planet $\varv_\mathrm{p}$ is expressed as
\begin{equation}
    \varv_{\mathrm{p}}=\varv_{\mathrm{sys}}-\varv_{\mathrm{bary}}+K_{\mathrm{p}} \sin (2 \pi \phi)+\Delta \varv,
\end{equation}
where $\varv_\mathrm{sys}$ is the systemic velocity, $\varv_\mathrm{bary}$ is the barycentric Earth radial velocity (BERV), $\Delta \varv$ is the radial velocity deviation from planetary rest frame, and $\phi$ is the orbital phase. A phase-folded one-dimensional CCF was generated for each assumed $K_\mathrm{p}$ value by averaging the in-transit CCF map after being shifted to the planetary rest frame by that specific $K_\mathrm{p}$. A $K_\mathrm{p}-\Delta \varv$ map was then created by stacking the phase-folded one-dimensional CCFs corresponding to different $K_\mathrm{p}$ values, ranging from 0 to 300 km~s$^{-1}$ with a step of 3 km~s$^{-1}$. The $K_\mathrm{p}-\Delta \varv$ map was normalized by the standard deviation calculated from $+$0 to $+$120 km~s$^{-1}$ in $K_\mathrm{p}$ and $+$30 to $+$150 km~s$^{-1}$ in $\Delta \varv$. Consequently, the final $K_\mathrm{p}-\Delta \varv$ map was expressed in the form of S/N.

\section{Results and Discussion}\label{RandD}

\subsection{Evidence for H$_2$O Emission During WASP-33b's Transit}\label{WASP33bNightInversion}

We employed the transmission spectrum models that incorporate planetary nightside thermal emission (as detailed in Section~\ref{CRMofNPA}) to conduct a CCF analysis for WASP-33b using data from both CARMENES and GIANO-B (see Table~\ref{OBSINFO} for the observational logs). We set up different T-P profiles at the limb and nightside to compute $R_\mathrm{eff}$ and $F_\mathrm{p}$, respectively.

\begin{figure}
\centering
\includegraphics[width=\hsize]{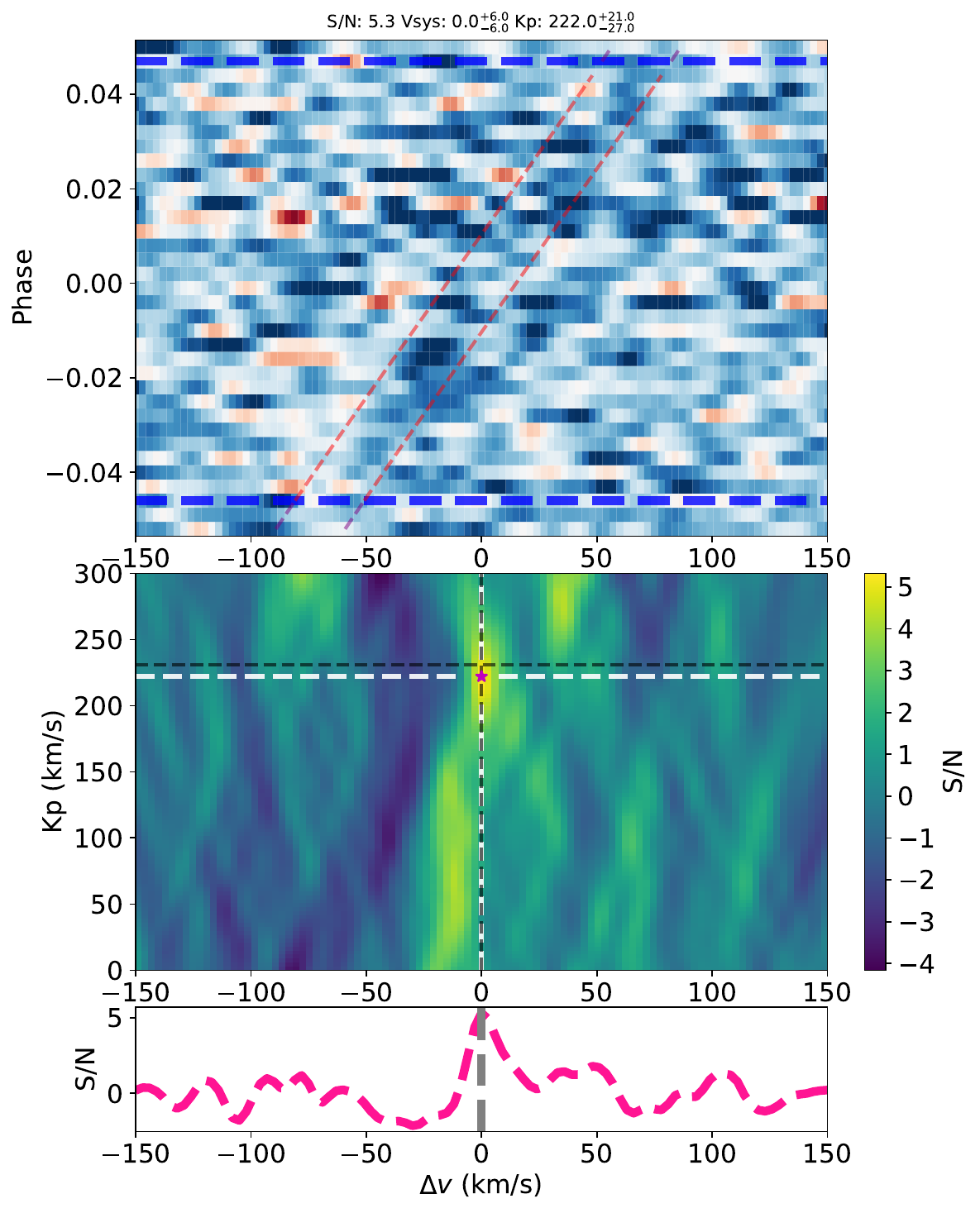}
\caption{CCF and $K_\mathrm{p}-\Delta \varv$ maps of H$_2$O averaged over two nights of GIANO-B observations. First row: CCF map, where the two horizontal dashed lines indicate the start and end of the transit, and the tilted dashed line represents the expected planetary orbit. Second row: $K_\mathrm{p}-\Delta \varv$ map with orthogonal black dashed lines for the expected $K_\mathrm{p}$ and $\Delta \varv$, and a pentagram for the position of the S/N peak. Third row: the one-dimensional CCF corresponding to the $K_\mathrm{p}$ value at the S/N peak. The vertical gray dashed line indicates $\Delta \varv \text{ = } 0$.}
\label{Fig2}%
\end{figure}

We found evidence for the H$_2$O emission features during the transit of WASP-33b. The transmission spectrum model we used is shown in Figure~\ref{Fig1}. The H$_2$O emission signal is detected in the two GIANO-B transit observations with S/N $\text{= } 4.0$ and S/N $\text{= } 5.5$, respectively. By combining the observations from the two nights, we obtained the final $K_\mathrm{p}-\Delta \varv$ map shown in Figure~\ref{Fig2}. The H$_2$O peak signal is located at $K_\mathrm{p} \text{ = } 222_{-27}^{+21} \text{ km~s}^{-1}$ and $\Delta \varv \text{ = } 0_{-6}^{+6} \text{ km~s}^{-1}$ with an S/N of 5.3. This $K_\mathrm{p}$ is consistent with the expected $K_\mathrm{p} \text{ = } 231 \text{ km~s}^{-1}$ calculated from the orbital parameters of WASP-33b. A near-zero value for $\Delta \varv$ suggests that there is no evidence for a strong day-night wind and/or an equatorial super-rotation jet in WASP-33b from the current observational data. 

Previous optical transmission spectroscopy to detect atoms and ions in WASP-33b showed no significant circulation patterns \citep{2024AJ....167...36Y,2021A&A...645A..22Y}, consistent with the results here. However, due to the strong influence of stellar pulsations, caution should be bared in mind when claiming species detection from the transmission spectrum of WASP-33b  \citep{2024AJ....167...36Y,2021A&A...645A..22Y,2021AJ....161..152C,2021A&A...653A.104B}. Fortunately, H$_2$O will not be present in WASP-33b's host star due to the extremely high effective temperatures of A-type stars, so the effect of pulsations on the detection of H$_2$O emission signals from WASP-33b should be negligible. 

The observed S/N peaks of the H$_2$O emission signal from the two nights are consistent at the 1.5$\sigma$ level (see~Appendix~\ref{Appendix_KPMAP}). The slight deviation could be due to insufficient data quality and imperfect data reduction, or could indicate the presence of planetary variability and a particular planetary climate pattern similar to WASP-121b \citep{2024ApJS..270...34C}. On the other hand, we did not find robust emission signals from H$_2$O in the CARMENES data, which we attribute to the insufficient coverage of the infrared band for CARMENES. We tested truncating the GIANO-B data at $\lesssim$1.75 $\micron$ to match the wavelength coverage of CARMENES and found that the detected emission signature would almost disappear (see Appendix~\ref{truncating_giano}).

\subsection{Does WASP-33b's Nightside Have an Inverted Temperature Structure?}\label{WASP33bNightInversionInterpretation}
The inference of an emission signal indicates the existence of an inverted temperature structure on the nightside. According to the general understanding of hot Jupiters from theoretical models, the temperature difference between the dayside and the nightside essentially depends on the nature of the circulation, which is roughly determined by comparing the radiative timescale ($\tau_\mathrm{rad}$) with the wave propagation timescale or the advection timescale \citep[$\tau_\mathrm{wave}$ or $\tau_\mathrm{adv}$,][]{2002A&A...385..166S,2016ApJ...821...16K}. In the equatorial region, there is no Coriolis force, with advection balancing the day-night pressure gradient force, and additional consideration of the Coriolis force is required at mid and high latitudes. As a result, as the planetary equilibrium temperature increases, $\tau_\mathrm{rad}$ becomes shorter, leading to an increased day-night temperature difference. Such a pattern has already been confirmed by phase curve observations of hot Jupiters \citep{2020RAA....20...99Z}. 

However, compared to classical hot Jupiters, UHJs are subject to dissociation of molecules and ionization of atoms on the dayside due to higher dayside temperatures, and the radiative feedback from these processes affects the day-night heat transport, i.e. provides an additional energy budget. Several studies \citep{2018ApJ...857L..20B,2018RNAAS...2...36K,2019ApJ...886...26T} have suggested that the H$_2$-H dissociation and recombination significantly increase the day-night heat transport, reducing the day-night temperature difference and the velocity of the equatorial super-rotation jets. In this process, H$_2$ dissociates into H on the dayside, which cools the atmosphere; the H produced by the dissociation on the dayside is transported by the circulation to recombine into H$_2$ on the nightside, which heats the atmosphere. If our H$_2$O emission inference is correct, this new energy budget will make the nightside of the planet hotter than previously expected by the GCM model, indicating more efficient energy transport.

The presence of nightside thermal inversions seems to be an overheating of the upper atmosphere driven by atmospheric circulation. The mass mixing ratio of atomic hydrogen relative to the total gas, $q = \rho_\mathrm{H} /(\rho_\mathrm{H} + \rho_{\mathrm{H}_2})$, increases with decreasing pressure, therefore corresponds to stronger horizontal heat transport at lower pressures. Meanwhile, the radiative cooling efficiency likely does not increase with decreasing pressure as fast as the heat transport. These two processes help to maintain a thermal inversion at part of the nightside. Due to the extremely short timescale of the H recombination process, chemical heating from this process occurs predominantly near the terminator of UHJs, as predicted by current state-of-the-art models.

Here, however, we find evidence for nightside thermal inversion, suggesting that more energy can be transported closer to the anti-stellar point. \citet{2024MNRAS.528.1016T} recently suggested that the formation of a nightside thermal inversion layer is possible when considering H$_2$-H dissociation and recombination in planets with equilibrium temperatures above 2600 K and the absence of radiative cooling due to the lack of nightside coolants (e.g., TiO, VO). TiO has been detected on the dayside of WASP-33b by emission spectroscopy \citep{2021A&A...651A..33C}, but not on the terminator by transmission spectroscopy \citep{2024AJ....167...36Y}. This implies that the condensation of TiO occurs on the terminator/nightside and could then be circulated to the dayside to restore the gas phase. In addition, the heat transport and 3D thermal structure of UHJs are affected by other factors such as drag/damping processes (especially magnetic drag) and radiative feedback from nightside clouds. 

An inverted temperature structure on the nightside is closely related to the nature of the atmospheric circulation, the efficiency of heat transport, and the specific energy budget. However, the 1D model ignores the effects of the 3D atmospheric structure and radiative transfer, as shown in Equation~(\ref{DeltaF}) in Appendix~\ref{DGE}. The factor that dominates these effects is the rotation of the visible hemisphere during the observation. A portion of the limb and dayside atmosphere has already rotated into the visible hemisphere during the transit and is not fully visible on the nightside due to planetary rotation. We conducted a phase-resolved high-resolution simulation of H$_2$O thermal emission spectra incorporating 3D velocity fields with PICASO \citep{2019ApJ...878...70B,2023ApJ...942...71M} by post-processing the GCM outputs (see Appendix~\ref{HRPCS}). This simulation indicates that the 3D effects can cause the planetary thermal emission during transit to exhibit emission features for a planet as hot as WASP-33b (see Figure~\ref{HRPCS_Figure}). As a result, the coupled transmission spectrum appears emission-like. This simulation supports our alternative speculation that the limb and dayside inverted temperature structures may also contribute to the emission features we detected.

In summary, we attribute the detected emission features to the nightside thermal inversion driven by atmospheric dynamics or the 3D effects. To fully interpret the detected emission features, it is necessary to consider the 3D atmospheric structure and radiative transfer effects to determine the precise boundary of the thermal inversion region. Further observations and the development of theories and models to better constrain the 3D chemistry, circulation, and thermal structure of UHJs are underway. WASP-33b is one of the best targets for testing these theories. Further quantitative investigations will be explored in a future paper.

\begin{figure*}
\centering
\includegraphics[width=\hsize]{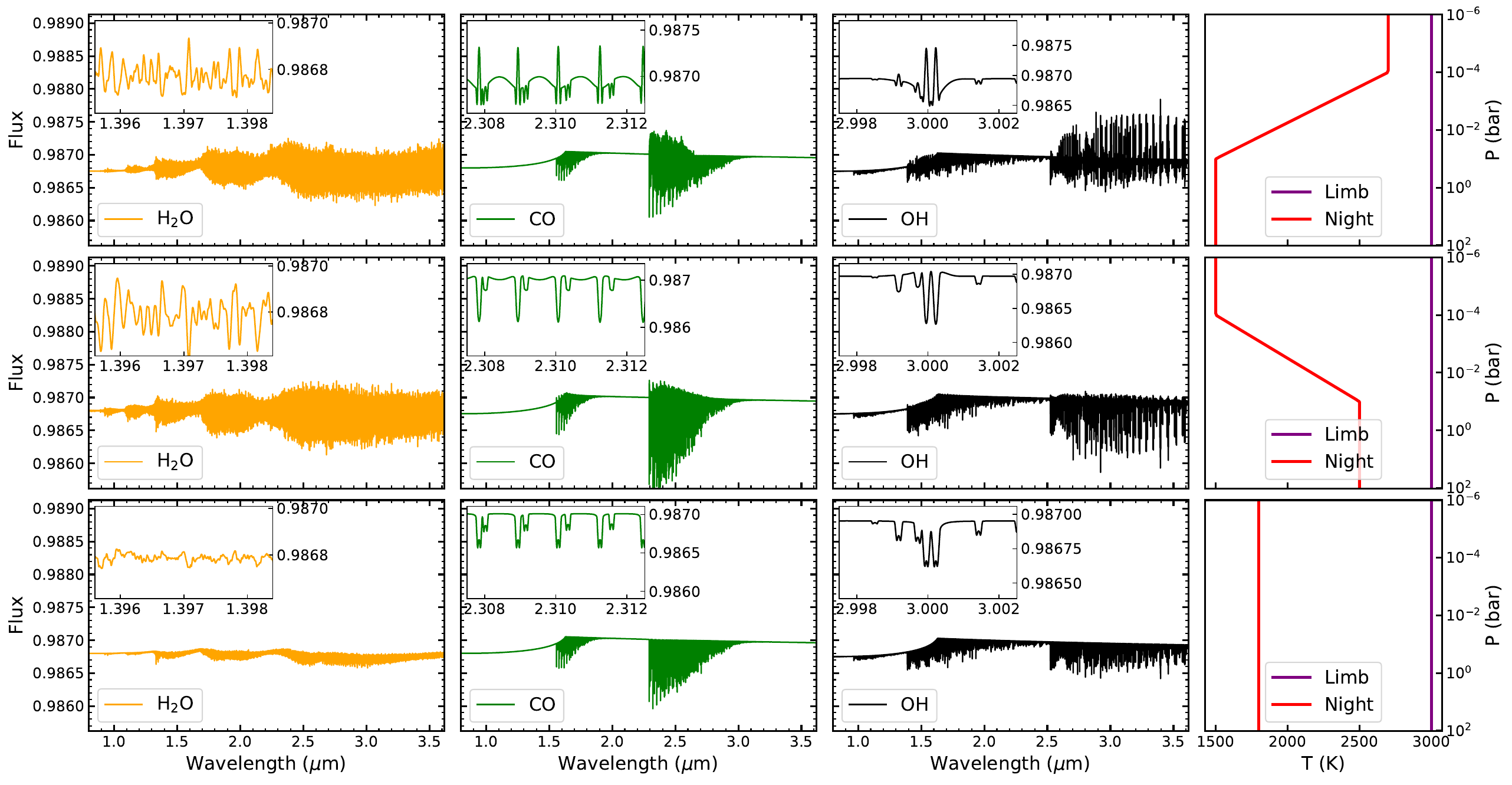}
\caption{The general transmission spectrum models with three types of nightside T-P profiles for WASP-33b. The first three columns show the general transmission spectra of H$_2$O, CO, and OH, respectively, while the fourth column shows the T-P profiles. Each inset shows a zoomed view of the spectrum. From top to bottom correspond to the inverted, non-inverted, and isothermal nightside T-P profiles.}
\label{Fig3}%
\end{figure*}

\subsection{Effects of Nightside Thermal Emission on Transmission Spectrum Observations}
While the dilution effect of planetary nightside emission on transmission spectra has been discussed in several low-resolution observational studies \citep[][]{2014AJ....147..161S,2018AJ....156...17K,2021AJ....162...34K}, it has never been considered in the high-resolution transmission spectroscopy observations. We show that an approximation using the linear source functions at local thermodynamic equilibrium in Equations~(\ref{MEA1})-(\ref{MEA3}) in Appendix~\ref{DGE}. It follows that the effect of thermal emission on the transmission spectrum is determined by the nightside vertical temperature gradient and the ratio of the line and continuum absorption coefficients. That is, the larger the temperature gradient, line cross-section, and species abundance, the more significant the nightside thermal emission. 

According to Section~\ref{RM}, we set up three types of temperature structures for the nightside of WASP-33b: (a) an inverted profile, (b) a non-inverted profile, and (c) an isothermal profile. For the limb, we simply assume an isothermal profile, although the limb temperature structure of the UHJ is generally inverted according to advanced GCM predictions, but the variation of the limb temperature structure is more a modulation of the atmospheric scale height and has a limited effect on the spectral line shape. 

Figure~\ref{Fig3} shows the transmission spectrum models with the three types of nightside temperature structures for H$_2$O, CO and OH. We find that when considering the nightside thermal emission of WASP-33b, whenever the nightside is non-isothermal, it affects the transmission spectra of the molecules to different degrees. When the nightside has an inverted temperature structure, the H$_2$O line shows pure emission features, while the other two molecules show absorption line wings and emission line cores, which weaken the transmission spectral signals, making the detection of these two molecules difficult. In contrast, the thermal emission produced by the non-inverted nightside temperature structure enhances the transmission spectral features of the three molecules. 

In addition to these three molecules, the nightside thermal emission should also affect the transmission spectra of other molecules. It depends on the coincidence between the excitation temperatures of the spectral lines and the different temperature structures corresponding to different planets. We emphasize that the nightside thermal emission of UHJs should not be neglected in high-resolution infrared transit observations, and that at least some tests are required to assess the effects of the nightside thermal emission, especially in the case of non-isothermal nightside temperature structures. 

\begin{figure}
\centering
\includegraphics[width=\hsize]{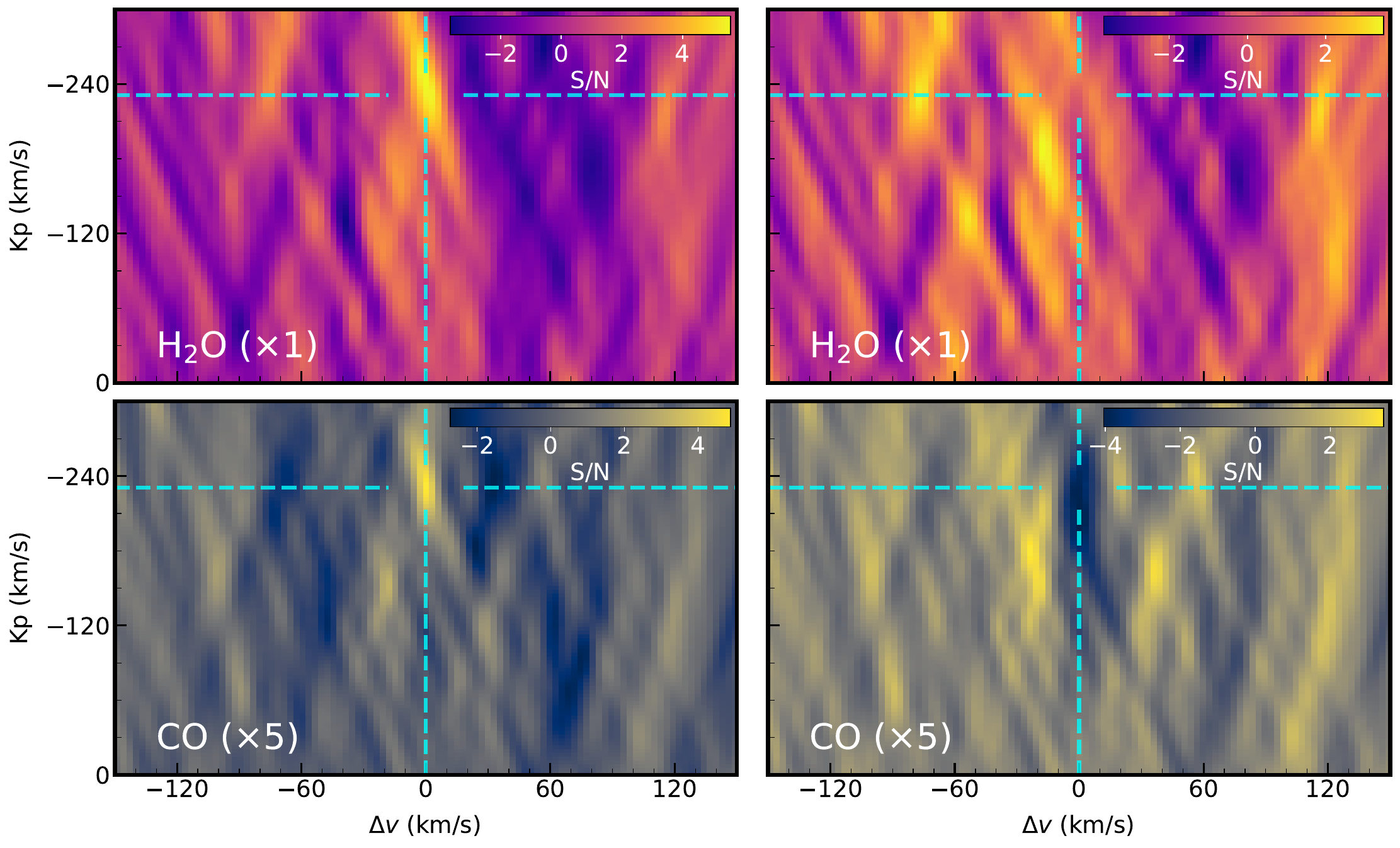}
\caption{Comparison of $K_\mathrm{p}-\Delta \varv$ maps for the injection-recovery tests of WASP-33b. The injected general transmission model has included the nightside thermal emission. The cyan dashed lines indicate the expected values of the injected signal. The left and right panels show the results from the correct general transmission spectral template and the pure transmission spectral template, respectively. The first and second rows present the results for H$_2$O and CO, respectively. The strength of the injected CO signal has been enhanced by five times. 
}
\label{Fig4}%
\end{figure}	

Certainly, a non-negligible problem is the obscuring effect of clouds, but in the case of some hot Jupiters (UHJs or planets with high heat transport efficiency) the nightside temperatures are likely to be higher than the condensation temperatures of the various condensates, and the effect of the nightside thermal emission should not be neglected. Current GCM simulations show that the nightside cloud cover of UHJs is partial and dynamic, which is significantly different from the uniform cloud cover of hot Jupiters \citep{2022ApJ...934...79K}. Moreover, the mean cloud top pressures on the nightside are $\sim$10–100 mbar, which is deeper than the temperature inversion caused by the thermal effects of hydrogen recombination at $\sim$1–10 mbar in GCMs. Therefore, the presence of these clouds would not affect the emission lines.

To illustrate how molecular detection would be affected if the nightside thermal emission of UHJs were ignored, we performed an injection-recovery test using H$_2$O and CO as an example. The general transmission model created in the top panel of Figure~\ref{Fig3} was injected into the observational data, and then a cross-correlation analysis was performed with a general transmission spectral template with nightside emission and a pure transmission spectral template without nightside emission, separately. The results are shown in Figure~\ref{Fig4}. 

For H$_2$O, a pure transmission spectral template neglecting the nightside thermal emission would lead to a non-detection. The H$_2$O signal can only be recovered with the general transmission spectral template that includes the nightside thermal emission. For CO, the S/N of the current data cannot recover the injected signal even with the input general transmission model. It can be recovered only when the strength of the injected signal is increased five times. We note that neglecting the nightside thermal emission can lead to diametrically opposite features in the $K_\mathrm{p}-\Delta \varv$ map if the S/N is high enough or the signal strength is strong enough, which is due to the difference in the CO spectral line profiles between the general transmission model and the pure transmission model. 

In summary, the nightside thermal emission could have a significant impact on the molecular detection of UHJs, and the nightside temperature structure and cloud shielding effect need to be further investigated. If clouds do not completely obscure the nightside thermal emission, the nightside thermal and chemical properties can be inferred directly from the transmission spectra using Equation~(\ref{MainEqu}). 

\section{Conclusions}\label{CONCL}
	
We explore the influence of nightside thermal emission on high-resolution infrared transmission spectroscopy, especially in the case of UHJs. We present a general equation for the high-resolution transmission spectrum that includes planetary nightside thermal emission. This provides a new way to infer the thermal structure of the planetary nightside from high-resolution transmission spectroscopy observations. We find evidence for H$_2$O emission in the UHJ WASP-33b during transit with the GIANO-B archival data, and interpret it as the presence of a nightside temperature inversion with a 1D general high-resolution transmission spectrum model. The S/N of the signal is not high enough to claim a robust detection, and further high-quality observations are needed to confirm our results. 

We propose that some UHJs may exhibit non-isothermal temperature structures on the nightside, especially thermal inversions in the upper atmosphere due to higher than expected heat transport efficiencies, which would allow the nightside emission to penetrate the cloud layer. On the other hand, it is also possible that the 3D effect of the rotating visible hemisphere could cause a portion of the limb and dayside atmosphere to rotate into view during transit, resulting in emission features. Therefore, while it is necessary to consider the nightside thermal emission in high-resolution infrared transmission spectroscopy, a more in-depth study of the 3D chemical, circulation, and thermal structure of UHJs is also needed to fully understand the origin of a potential nightside thermal inversion.

\begin{acknowledgments}

The authors thank the anonymous reviewer for the constructive comments and suggestions on the manuscript. We thank Natasha Batalha for sharing the line-by-line opacity database used in computing the synthetic high-resolution spectrum in this work. G.C. acknowledges the support by the National Natural Science Foundation of China (NSFC; grant Nos. 12122308, 42075122), Youth Innovation Promotion Association CAS (2021315), and the Minor Planet Foundation of the Purple Mountain Observatory. F.Y. acknowledges the support by the NSFC (grant No. 42375118). J.J. acknowledges the support by the NSFC (grant No. 12033010).

\end{acknowledgments}

\bibliography{ref.bib}{}
\bibliographystyle{aasjournal}




\appendix

\section{Derivation of a Time-Resolved General Equation for Exoplanet High-Resolution Transmission Spectra.}\label{DGE}

Here we derive the general equation for high-resolution transmission spectrum that incorporates planetary nightside thermal emission shown in Equation~(\ref{MainEqu}) and simplify the equation to illustrate the effect of planetary nightside emission on the transmission spectrum. First we derive a formal equation for the time-resolved high-resolution transmission spectra for 3D atmospheric and radiative transfer calculations. We then focus on the case of the 1D atmospheric solution and present an approximate solution for a special situation. This derivation builds on the concept of \citet{2022ApJ...929...20M} for the low-resolution transmission spectrum.

The transmission spectrum is defined as the spectral flux difference between the out-of-transit time and the in-transit time. At high resolution, the transmission spectrum will be Doppler-shifted at each time (or phase) due to the orbital motion of the planet. Therefore, the formal equation for high-resolution transmission spectra will be a time-dependent equation:
\begin{equation}\label{A1}
    \Delta_{\lambda}(t) \equiv \frac{\overline{\mathcal{F}}_\mathrm{\lambda, out}(t) - \mathcal{F}_\mathrm{\lambda, in}(t)}{\overline{\mathcal{F}}_\mathrm{\lambda, out}(t)}.
\end{equation}
To unpack each terms, we have 
\begin{eqnarray}\label{A2}
    &&\mathcal{F}_\mathrm{\lambda, out}(t) \equiv \mathcal{F}_\mathrm{\lambda, s, out}(t) + \mathcal{F}_\mathrm{\lambda, p(night), out}(t), \nonumber \\
    &&\mathcal{F}_\mathrm{\lambda, in}(t) \equiv \mathcal{F}_\mathrm{\lambda, s, in}(t) + \mathcal{F}_\mathrm{\lambda, s, trans}(t) + \mathcal{F}_\mathrm{\lambda, p(night), in}(t),
\end{eqnarray}
where ``p(night)'' denotes the planetary nightside, ``s'' denotes the star, ``out'' denotes out-of-transit, ``in'' denotes in-transit, and ``trans'' denotes the transmission of stellar light through the planetary atmosphere. The average flux outside of transit can be expressed as
\begin{equation}\label{A3}
    \overline{\mathcal{F}}_\mathrm{\lambda, out}(t) = \mathcal{F}_\mathrm{\lambda, s, out} + \overline{\mathcal{F}}_\mathrm{\lambda, p(night), out}(t),
\end{equation}
where $\overline{\mathcal{F}}_\mathrm{\lambda, p(night), out}(t)$ is time-averaged planetary nightside thermal flux outside of transit. Substituting Equation~(\ref{A2}) and (\ref{A3}) into Equation~(\ref{A1}), we have
\begin{equation}\label{A4}
    \Delta_\mathrm{\lambda}(t) = \frac{\mathcal{F}_\mathrm{\lambda, s, out} - \mathcal{F}_\mathrm{\lambda, s, in}(t) - \mathcal{F}_\mathrm{\lambda, s, trans}(t) + \overline{\mathcal{F}}_\mathrm{\lambda, p(night), out}(t) - \mathcal{F}_\mathrm{\lambda, p(night), in}(t)}{\mathcal{F}_\mathrm{\lambda, s, out} + \overline{\mathcal{F}}_\mathrm{\lambda, p(night), out}(t)}.
\end{equation}
The flux is defined as,
\begin{equation}\label{A5}
    \mathcal{F}_{\lambda}=\int_{\Omega} I_{\lambda} \hat{\mathbf{n}} \cdot \hat{\mathbf{k}} d \Omega,
\end{equation}
where $I_\mathrm{\lambda}$ is the spectral intensity, $\Omega$ is the solid angle subtended by the source at the observer, and $\hat{\mathbf{n}}$ and $\hat{\mathbf{k}}$ are the unit vectors in the direction of the normal and observer, respectively. The stellar rays satisfy $\hat{\mathbf{n}} \cdot \hat{\mathbf{k}} = 1$ in the transit geometry, but the planetary thermal emission does not. Therefore, we have 
\begin{eqnarray}\label{A6}
    &\mathcal{F}_\mathrm{\lambda, s, out} &= \int_{\Omega_\mathrm{s(full)}} I_\mathrm{\lambda, s} d\Omega, \\
    &\mathcal{F}_\mathrm{\lambda, s, in}(t) &= \int_{\Omega_\mathrm{s, unobscured}(t)} I_\mathrm{\lambda, s} d\Omega, \\
    &\mathcal{F}_\mathrm{\lambda, s, trans}(t) &= \int_{\Omega_\mathrm{p}(t)} I_\mathrm{\lambda, s, trans}(t) d\Omega, \\
    &\mathcal{F}_\mathrm{\lambda, p(night)}(t) &= \oint V(\theta, \phi, t) I_\mathrm{\lambda, p}(\theta, \phi, t) d\Omega,
\end{eqnarray}
where the visibility function \citep[see][]{2018AJ....156..235R,2013MNRAS.434.2465C,2011ApJ...726...82C} is
\begin{equation}\label{A7}
    V(\theta, \phi, t) = \max \left[\cos \gamma_{0}, 0\right] = \max \left[\sin \theta \sin \theta_{0} \cos \left(\phi-\phi_{0}\right) + \cos \theta \cos \theta_{0}, 0\right],
\end{equation}
where $\theta$ and $\phi$ are the latitude and longitude, respectively, $\theta_{0}(t)=\theta_{0}$, $\phi_{0}(t)=\phi_{0}(0)-\omega_\mathrm{rot} t$, and $\omega_\mathrm{rot}$ is planetary rotation angular velocity. For tidally locked planets, $\omega_\mathrm{rot}=\omega_\mathrm{orbit}$. Moreover, we can rewrite 
\begin{eqnarray}\label{A8}
    &\mathcal{F}_\mathrm{\lambda, s, out} - \mathcal{F}_\mathrm{\lambda, s, in}(t) 
    &=\int_{\Omega_\mathrm{overlap}(t)} I_\mathrm{\lambda, s} d\Omega.
\end{eqnarray} 
Then, we introduce the mean ``stellar intensity'' and the time-averaged ``planetary thermal emission intensity'':
\begin{eqnarray}\label{A9}
    &\overline{I}_\mathrm{\lambda,s} &= \frac{\int_{\Omega_\mathrm{s(full)}} I_\mathrm{\lambda,s} d\Omega}{\int_{\Omega_\mathrm{s(full)}} d\Omega}, \nonumber \\
    &\overline{I}_\mathrm{\lambda,p(night)} &= \frac{\int \oint V(\theta, \phi, t) I_\mathrm{\lambda,p(night)}(\theta, \phi, t)d\Omega dt}{\int \oint V(\theta, \phi, t) d\Omega dt}.
\end{eqnarray}
Therefore, 
\begin{eqnarray}\label{A10}
    &\mathcal{F}_\mathrm{\lambda, s, out} &= \bar{I}_\mathrm{\lambda, s} \Omega_\mathrm{s(full)}, \nonumber \\
    &\overline{\mathcal{F}}_\mathrm{\lambda, p(night), out}(t) &= \overline{I}_\mathrm{\lambda, p(night)} \Omega_\mathrm{p(full)}.
\end{eqnarray}
Next, we replace the solid angle integral with the projected area integral. $\Omega_\mathrm{(full)} = A/D^2$, so $d \Omega_\mathrm{(full)} = dA/D^2$, where $A = \pi R^2$ is the projected area and $D$ is the distance from the star to the observer. Therefore, substituting Equation~(\ref{A6})-(\ref{A10}) into Equation~(\ref{A4}), we have
\begin{eqnarray}\label{A11}
    &\Delta_{\lambda}(t) = &\left[\int_{A_\mathrm{overlap}(t)} \frac{I_{\lambda, s}}{\overline{I}_{\lambda, s}} dA - \int_{A_\mathrm{p(full)}} \frac{I_\mathrm{\lambda, s, trans}(t)}{\overline{I}_{\lambda, s}} dA 
    + A_\mathrm{p(full)}\frac{\overline{I}_\mathrm{\lambda, p(night)}}{\overline{I}_\mathrm{\lambda, s}} - \int_{A_\mathrm{p(full)}} \frac{V(\theta, \phi, t) I_\mathrm{\lambda, p(night)}(\theta, \phi, t)}{\overline{I}_{\lambda, s}} dA \right] \nonumber \\ 
    &&\times \frac{1}{A_\mathrm{s(full)}+A_\mathrm{p(full)}\frac{\overline{I}_\mathrm{\lambda, p(night)}}{\overline{I}_\mathrm{\lambda, s}}}.
\end{eqnarray}
Next, we follow \citet{2022ApJ...929...20M} to reduce $I_\mathrm{\lambda, s, trans}$, we have
\begin{eqnarray}\label{A12}
    &I_\mathrm{\lambda, s, trans}(t) &= \sum_{m=1}^{N_\mathrm{phot}} \delta_\mathrm{ray,s,m} \mathcal{T}_\mathrm{\lambda, m}(t) I_\mathrm{\lambda, s, m} \nonumber \\
    &&=\overline{\delta_\mathrm{ray, s} \mathcal{T}_\mathrm{\lambda}(t)} I_\mathrm{\lambda, s},
\end{eqnarray}
where
\begin{equation}\label{A13}
    \delta_\mathrm{ray, s} = \left\{\begin{array}{ll}
    1, & \text{if ray intersects star.} \\
    0, & \text{else.}
    \end{array}\right.
\end{equation}
and 
\begin{eqnarray}\label{A14}
    &&\mathcal{T}_\mathrm{\lambda,m}(t)=e^{-\tau_\mathrm{\lambda, path, m}(t)}, \nonumber \\
    &&\tau_\mathrm{\lambda, path, m}(t)=\int_{0}^{\infty} \tilde{\alpha}_{\lambda}\left(s_{m}, t\right) d s_{m}.
\end{eqnarray}
Here, we neglect stellar limb-darkening, making $\frac{I_{\lambda, s}}{\overline{I}_{\lambda, s}}$ to be factored out of the integrals in Equation~(\ref{A11}). Therefore, 
\begin{eqnarray}\label{A15}
    &\Delta_{\lambda}(t) = &\left[
    \frac{I_\mathrm{\lambda, s}}{\overline{I}_\mathrm{\lambda, s}}A_\mathrm{overlap}(t) - \frac{I_\mathrm{\lambda, s}}{\overline{I}_\mathrm{\lambda, s}}\int_{A_\mathrm{p(full)}} \overline{\delta_\mathrm{ray, s} \mathcal{T}_{\lambda}(t)} dA + \frac{\overline{I}_\mathrm{\lambda, p(night)}}{\overline{I}_\mathrm{\lambda, s}}A_\mathrm{p(full)} - \int_{A_\mathrm{p(full)}} \frac{V(\theta, \phi, t) I_\mathrm{\lambda, p(night)}(\theta, \phi, t)}{\overline{I}_{\lambda, s}} dA \right] \nonumber\\
    &&\times \frac{1}{A_\mathrm{s(full)}+\frac{\overline{I}_\mathrm{\lambda, p(night)}}{\overline{I}_\mathrm{\lambda, s}}A_\mathrm{p(full)}}, 
\end{eqnarray}
where $A_\mathrm{overlap}$ denotes the overlapping area between the planetary and stellar disks \citep[see][]{2022ApJ...929...20M}. Then, we introduce two factors:
\begin{eqnarray}\label{A18}
    &\varepsilon_\mathrm{\lambda, het} &\equiv \frac{I_\mathrm{\lambda, s}}{\overline{I}_\mathrm{\lambda, s}} = \frac{1}{1-\sum_{i=1}^{N_\mathrm{het}} f_\mathrm{het, i}\left(1-\frac{I_\mathrm{\lambda, het, i}}{I_\mathrm{\lambda, s}}\right)}, \nonumber \\
    &\varepsilon_\mathrm{\lambda, night} &\equiv \frac{I_\mathrm{\lambda, p(night)}}{\overline{I}_\mathrm{\lambda, s}}. 
\end{eqnarray}
Therefore, substituting Equation~(\ref{A18}) into Equation~(\ref{A15}), we have 
\begin{equation}\label{DeltaF}
    \Delta_{\lambda}(t) = \left\{ \varepsilon_\mathrm{\lambda, het} \left[A_\mathrm{overlap}(t) - \int_{A_\mathrm{p}(t)} \overline{\delta_\mathrm{ray, s} \mathcal{T}_{\lambda}(t)} dA \right] + \overline{\varepsilon}_\mathrm{\lambda, night} A_\mathrm{p(full)} - \int_{A_\mathrm{p(full)}} \frac{V(\theta, \phi, t) I_\mathrm{\lambda, p(night)}(\theta, \phi, t)}{\bar{I}_{\lambda, s}} dA \right\} 
    \times \frac{1} {A_\mathrm{s(full)}+\overline{\varepsilon}_\mathrm{\lambda, night}A_\mathrm{p(full)}}.  
\end{equation}
The emergent planetary thermal emission intensity $I_\mathrm{\lambda, p(night)}$ can be solved by the radiative transfer equation (e.g., two-stream radiative transfer method). Its solution depends on the source function and optical depth, while the visible function depends on time, from which it follows that the flux depends strongly on time (orbital phase), reflecting the importance of the 3D atmospheric structure and radiative transfer calculations for transmission spectroscopy. 

Now let's start making approximations. The 3D effects are not the focus of this paper and we simplify to 1D atmospheres: 
\begin{eqnarray}\label{A20}
    &&I_\mathrm{\lambda, p(night)}(\theta, \phi, t) = I_\mathrm{\lambda, p(night)}(t), \nonumber \\
    &&V(\theta, \phi, t) = 1.
\end{eqnarray}
Therefore,
\begin{equation}\label{A21}
    \int_{A_\mathrm{p(full)}} \frac{V(\theta, \phi, t) I_\mathrm{\lambda, p(night)}(\theta, \phi, t)}{\overline{I}_{\lambda, s}} dA = \frac{I_\mathrm{\lambda, p(night)}(t)}{\overline{I}_\mathrm{\lambda, s}} \int_{A_\mathrm{p(full)}} dA = \varepsilon_\mathrm{\lambda, night}(t) A_\mathrm{p(full)}.
\end{equation}
Then, we introduce 
\begin{eqnarray}\label{A23}
    &&A_\mathrm{\lambda, eff}(t) \equiv A_\mathrm{overlap}(t)-\int_{A_\mathrm{p(full)}} \overline{\delta_\mathrm{ray, s} \mathcal{T}_{\lambda}(t)} d A = \pi R_\mathrm{\lambda, eff}^{2}(t), \nonumber \\
    &&d_\mathrm{\lambda, night} \equiv \frac{1}{1+\frac{\overline{\mathcal{F}}_\mathrm{\lambda, p(night), out}(t)}{\mathcal{F}_\mathrm{\lambda, s, out}}} = \frac{1}{1+\frac{A_\mathrm{p(full)} \overline{I}_\mathrm{\lambda, p(night)}}{A_\mathrm{s(full)} \overline{I}_\mathrm{\lambda, s}}} = \frac{A_\mathrm{s(full)}}{A_\mathrm{s(full)}+A_\mathrm{p(full)} \overline{\varepsilon}_\mathrm{\lambda, night}}.
\end{eqnarray}
Substituting Equation~(\ref{A21}) and (\ref{A23}) into Equation~(\ref{DeltaF}) and omitting ``full'', we have 
\begin{equation}
    \Delta_{\lambda}(t)=\left\{\varepsilon_\mathrm{\lambda, het} \frac{A_\mathrm{\lambda, eff}(t)}{A_\mathrm{s}}+\frac{A_\mathrm{p}}{A_\mathrm{s}}\left[\bar{\varepsilon}_\mathrm{\lambda, night}-\varepsilon_\mathrm{\lambda, night}(t)\right]\right\} \times d_\mathrm{\lambda, night}.
\end{equation}
Taking into account the photospheric radius correction \citep{2019ApJ...880L..16F}, $A_\mathrm{p} = A_\mathrm{\lambda, eff}(t)$, we have,
\begin{equation}\label{DeltaF1D}
    \Delta_{\lambda}(t)=\left\{\varepsilon_\mathrm{\lambda, het} \frac{A_\mathrm{\lambda, eff}(t)}{A_\mathrm{s}}+\frac{A_\mathrm{\lambda, eff}(t)}{A_\mathrm{s}}\left[\bar{\varepsilon}_\mathrm{\lambda, night}-\varepsilon_\mathrm{\lambda, night}(t)\right]\right\} \times d_\mathrm{\lambda, night}.
\end{equation}
Alternatively, we introduce an approximate line intensity $L$,
\begin{eqnarray}\label{A25}
    &L_\mathrm{\lambda, p(night)}(t) &\equiv I_\mathrm{\lambda, p(night)}(t) - \overline{I}_\mathrm{\lambda, p(night)} \nonumber \\
    &&\approx I_\mathrm{\lambda, p(night)}(t) - \overline{I}_\mathrm{\lambda, p(night), continuum}.
\end{eqnarray}
Substituting Equation~(\ref{A25}) into Equation~(\ref{DeltaF1D}), we have
\begin{equation}\label{De_MainEqu}
    \Delta_{\lambda}(t)=\left[\varepsilon_\mathrm{\lambda, het} \frac{A_\mathrm{\lambda, eff}(t)}{A_\mathrm{s}}-\frac{L_\mathrm{\lambda, p(night)}(t)}{\overline{I}_\mathrm{\lambda, s}} \frac{A_\mathrm{\lambda, eff}(t)}{A_\mathrm{s}}\right] \times d_\mathrm{\lambda, night},
\end{equation}
which completes the derivation of Equation~(\ref{MainEqu}), our unified equation for high-resolution transmission spectra with 1D atmosphere.

Next, we further analyze this equation in order to discuss what factors influence the intensity of the transmission spectrum. The formal solution for 1D plane-parallel radiative transfer equation (without scattering and $\mu = 1$) is
\begin{equation}\label{I}
    I_\mathrm{\lambda} = I_{\lambda}^\mathrm{bot} e^{-\tau_\mathrm{\lambda, vert}^{\text {bot }}}+\int_{0}^{\tau_\mathrm{\lambda, vert}^{\text {bot }}} S_\lambda(\tau_\mathrm{\lambda, vert}) e^{-\tau_\mathrm{\lambda, vert}} d \tau_\mathrm{\lambda, vert},
\end{equation}
where $S_\mathrm{\lambda}(\tau)$ is source function, $\tau$ is optical depth. Similarly, we introduce
\begin{equation}\label{Tn}
    \mathcal{T}_\mathrm{\lambda, vert} \equiv e^{-\tau_\mathrm{\lambda, vert}},
\end{equation}
which denotes the atmospheric transmission from a given layer to the top of the atmosphere. Substituting Equation~(\ref{Tn}) into Equation~(\ref{I}), we have 
\begin{equation}\label{FS1dRT}
    I_\mathrm{\lambda}(\mu) = I_\mathrm{\lambda}^\mathrm{bot} \mathcal{T}_\mathrm{\lambda, vert, bot} + \int_{\mathcal{T}_{\lambda, \mathrm{vert, bot}}}^{1} S_\mathrm{\lambda}(\mathcal{T}_\mathrm{\lambda, vert}) d \mathcal{T}_\mathrm{\lambda, vert},
\end{equation}
where the core is to solve for $S_\mathrm{\lambda}$.

To show $L$, next, we introduce Milne-Eddington approximation,  
\begin{equation}\label{MEA1}
    \tau_{\lambda}=\int \frac{k_{l}+k_{c}}{k_{c}} d \tau_{c},
\end{equation}
where 
\begin{equation}\label{MEA2}
    \frac{k_{l}}{k_{c}} = \eta_\mathrm{\lambda} \mathrm{(constant)}, \quad S\left(\tau_{\lambda}\right)=S_\mathrm{0}+\frac{d S}{d \tau_{c}} \tau_{c},
\end{equation}
where $k_{l}$ and $k_\mathrm{c}$ are line absorption coefficient and continuum absorption coefficient, respectively. In local thermodynamic equilibrium, $S = B(T_\mathrm{vert})$.

Substituting Equation~(\ref{FS1dRT})-(\ref{MEA2}) into Equation~(\ref{A25}), therefore, 
\begin{equation}\label{MEA3}
L = \frac{d B(T_\mathrm{vert})}{d \tau_{c}} \cdot \left(\frac{1}{1+\eta_{\lambda}}-1\right).
\end{equation}

\section{A Detailed Description of Data Reduction}\label{Appendix_DR}

The observation logs for the archival data we used are shown in Table~\ref{OBSINFO}. The raw CARMENES data were reduced using the CARACAL pipeline \citep{2016SPIE.9910E..0EC}. The GIANO-B observations were performed in ABAB nodding mode, which allows efficient removal of sky background and detector noise.	The raw spectra were reduced using the GOFIO pipeline \citep{2018SPIE10702E..66R}. The pipeline produces three spectra, including nodding A, nodding B, and combined nodding AB. In this work, only the AB spectra are used. Although the GOFIO pipeline performed a preliminary wavelength calibration using the U-Ne lamp spectra as a template, the mechanical instability of GIANO-B made the wavelength solution unstable in the observations. To prevent saturation of some emission lines from contaminating the observations, the U-Ne lamp spectra were acquired only at the end of the observations, resulting in an inaccurate wavelength solution determined by GOFIO \citep{2018A&A...615A..16B}. Therefore, we realigned the spectra and performed wavelength corrections using telluric lines following the approach of \citet{2019A&A...625A.107G}, \citet{2021Natur.592..205G}, and \citet{2022A&A...661L...6Y}. After raw data reduction by the pipelines, both CARMENES and GIANO-B data are archived as order-by-order spectra at the vacuum wavelength and in the terrestrial rest frame.

\begin{table*}[htb!]
\caption{Summary of the Observation Logs.}         
\label{OBSINFO}      
\centering          
\begin{tabular}{l c c c c c c}     
\hline\hline 
\noalign{\smallskip}      
Instrument & Target & Night & Date [UT] & Exposure Time [s] & $N_\mathrm{obs}$  & Airmass\\ 
\noalign{\smallskip} 
\hline\noalign{\smallskip} 
    GIANO-B & WASP-33b & Night-G1 & 2018-10-17 & 200& 167 & 1.66-1.01-1.49\\
    GIANO-B & WASP-33b & Night-G2 & 2018-11-08 & 200& 190 & 1.77-1.01-1.68\\ 
\hline\noalign{\smallskip} 
    CARMENES & WASP-33b & Night-C1 & 2017-01-05 & 120& 94 & 1.00-1.00-1.52\\
    CARMENES & WASP-33b & Night-C2 & 2018-11-30 & 250& 31 & 1.15-1.00-1.08\\ 
    CARMENES & WASP-33b & Night-C3 & 2018-12-05 & 200& 64 & 1.39-1.00-1.14\\
    CARMENES & WASP-33b & Night-C4 & 2019-10-01 & 120& 101 & 2.04-1.15-1.00\\
    CARMENES & WASP-33b & Night-C5 & 2019-11-03 & 210& 45 & 1.75-1.26-1.06\\
\noalign{\smallskip}
\hline                  
\end{tabular}
\end{table*}

\begin{figure}
\centering
\includegraphics[trim=0cm 0cm 0.5cm 0cm, clip=true,width=\hsize]{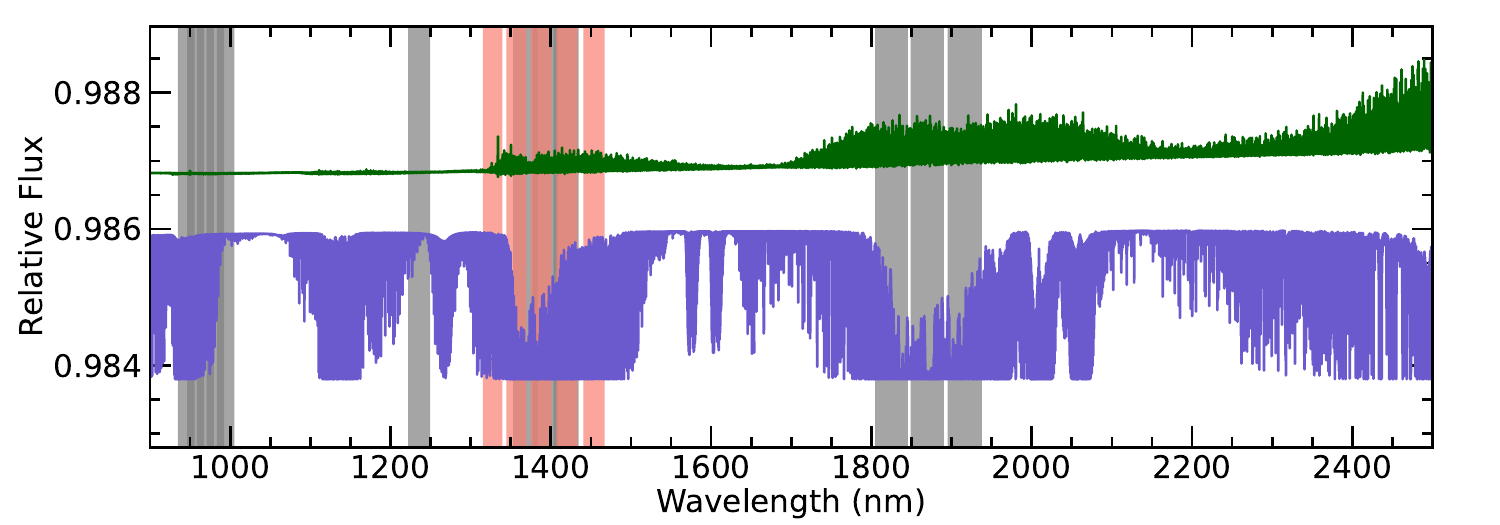}
\caption{Masking of the CARMENES and GIANO-B spectral orders. The vertical red and gray shaded areas indicate the masked orders for CARMENES and GIANO-B, respectively. The green line indicates the H$_2$O thermal emission spectrum. The purple line indicates the Earth's telluric spectrum.}
\label{FigMASK}
\end{figure}

We adopt the method detailed by \citet{2022A&A...659A...7Y} and \citet{2024AJ....167...36Y} to perform the subsequent spectral data reduction. To correct for outliers that are unlikely to be caused by the observed source, we flagged the pixels that were greater than five standard deviations from the flux values of the smoothed spectrum produced by a median filter with a window size of nine pixels for each order. The flux values of these pixels were replaced by the median of the thirty surrounding pixels \citep{2021MNRAS.502.4392L}. To correct for continuum variation, we divided each spectrum by the best-fit polynomial function, which is derived by fitting the ratio of each individual spectrum to the reference spectrum (i.e., the average of all spectra) with a fourth-order polynomial function \citep{2020A&A...635A.171C,2024AJ....167...36Y}. We constructed the spectral matrix by collecting all spectra of each observation, sorted by time. For GIANO-B, we discarded orders 8, 9, 10, 23, 24, 30, 45, 46, 47, 48, and 49 from our analysis, where order 0 is the reddest and order 49 is the bluest, due to saturated telluric absorption at certain wavelengths or few telluric lines to obtain wavelength corrections, similar to \citet{2021Natur.592..205G}. For CARMENES, we discard orders 6, 7, 8, 9, and 10, where order 0 is the reddest and order 27 is the bluest. The discarded orders are shown in Figure~\ref{FigMASK}.

To remove strong telluric regions, we masked the deep telluric regions that fall below 30\% of the median of the master spectrum, which is calculated by averaging all spectra for each order and contains the region around the deep telluric lines below 95\% of the median of the master spectrum. To remove strong sky emission lines, we masked the region above 150\% of the median of the continuum normalized master spectrum (containing the region above 105\% around each sky emission line). In addition, we further removed the 4$\sigma$ outliers in each wavelength channel (i.e., each column) using an iterative mode. Finally, we normalized each spectrum with a Gaussian high-pass filter. We used the SYSREM algorithm \citep{2005MNRAS.356.1466T} to reduce the telluric and stellar lines in the terrestrial rest frame. The SYSREM algorithm is used to fit each wavelength channel (each column) and each spectrum (each row) in the spectral matrix to capture systematics due to telluric contamination, stellar lines, and instrumental effects. These systematics are then removed from the spectral matrix. The SYSREM iteratively refines this process, yielding residual spectral matrices that retain planetary features and noise. We followed the approach proposed by \citet{2020MNRAS.493.2215G}, which divides the spectral matrix by the systematics rather than subtracting them. We ran the algorithm for fifteen iterations and selected the residual spectral matrix that yielded the maximum detection significance as the final output.

\section{Individual $K_\mathrm{p}-\Delta \varv$ map for each night}\label{Appendix_KPMAP}

\begin{figure}
\centering
\includegraphics[width=0.45\hsize]{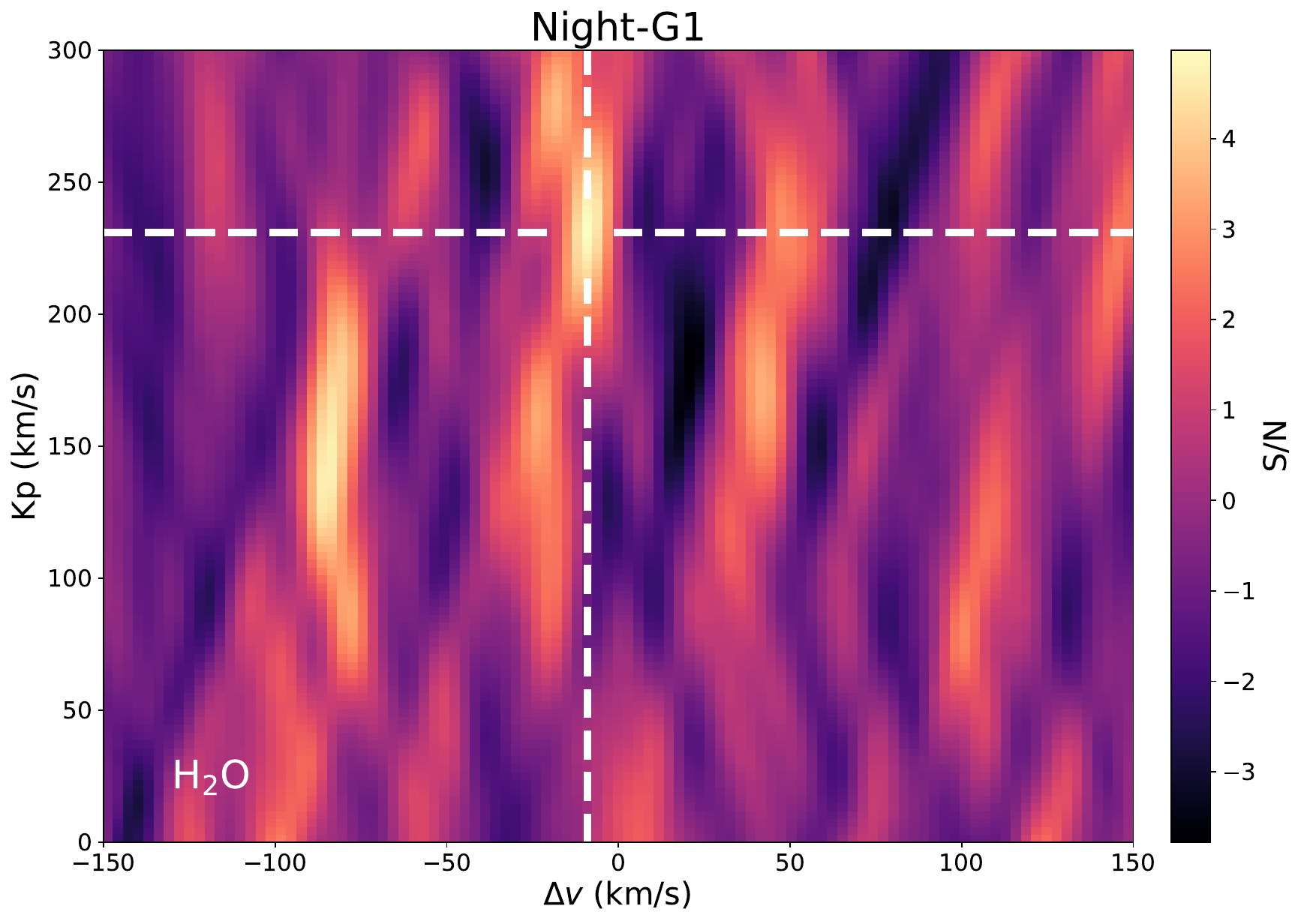}
\includegraphics[width=0.45\hsize]{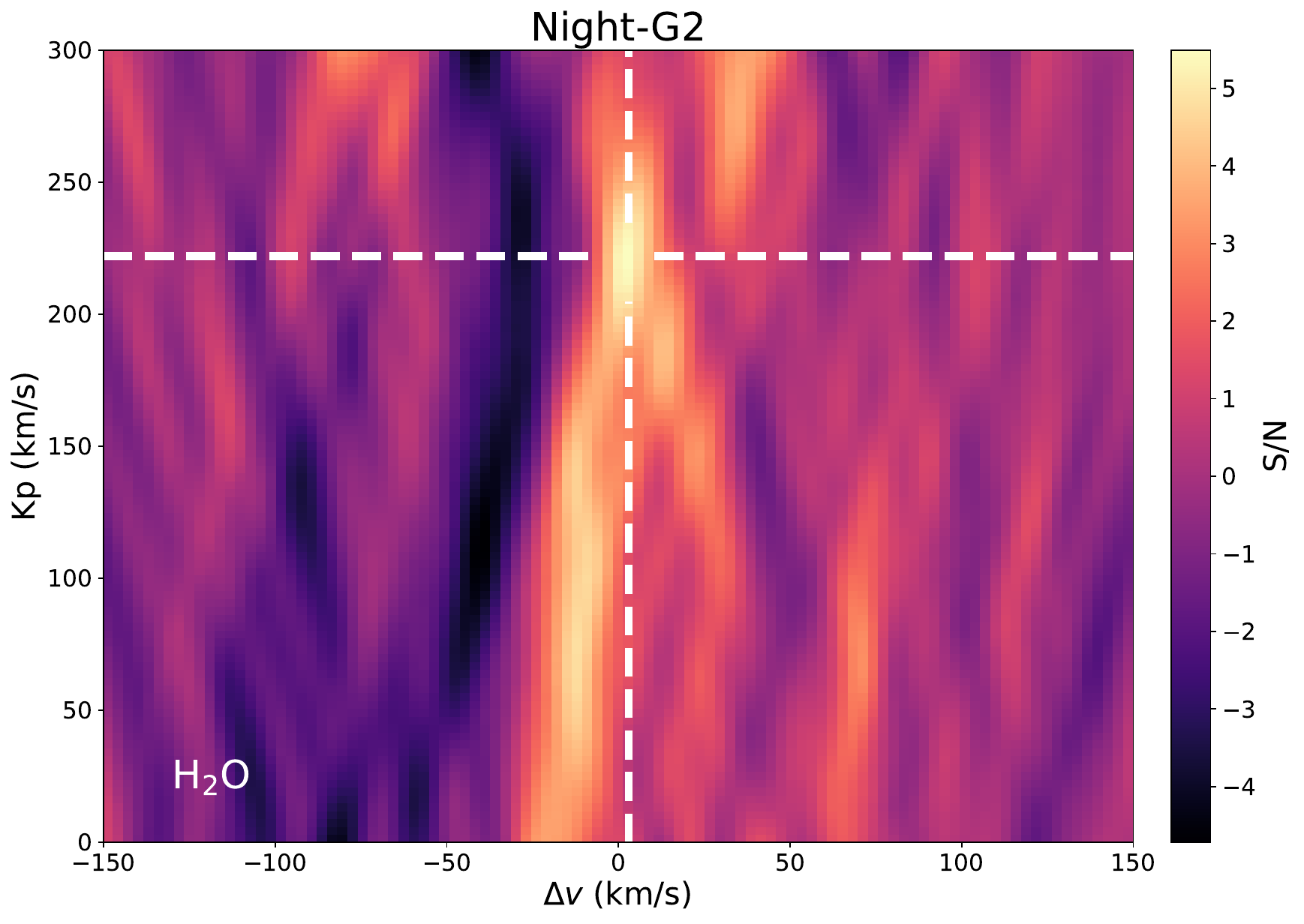}
\caption{$K_\mathrm{p}-\Delta \varv$ maps of H$_2$O from individual nights of GIANO-B observations. The crossed white dashed lines indicate the S/N peak.}
\label{INV_KPMAP}
\end{figure}

The individual $K_\mathrm{p}-\Delta \varv$ maps for each GIANO-B observation are shown in Figure~\ref{INV_KPMAP}. 

\section{Truncated GIANO-B data to resemble the insufficient wavelength coverage of CARMENES}\label{truncating_giano}

\begin{figure}
\centering
\includegraphics[width=0.45\hsize]{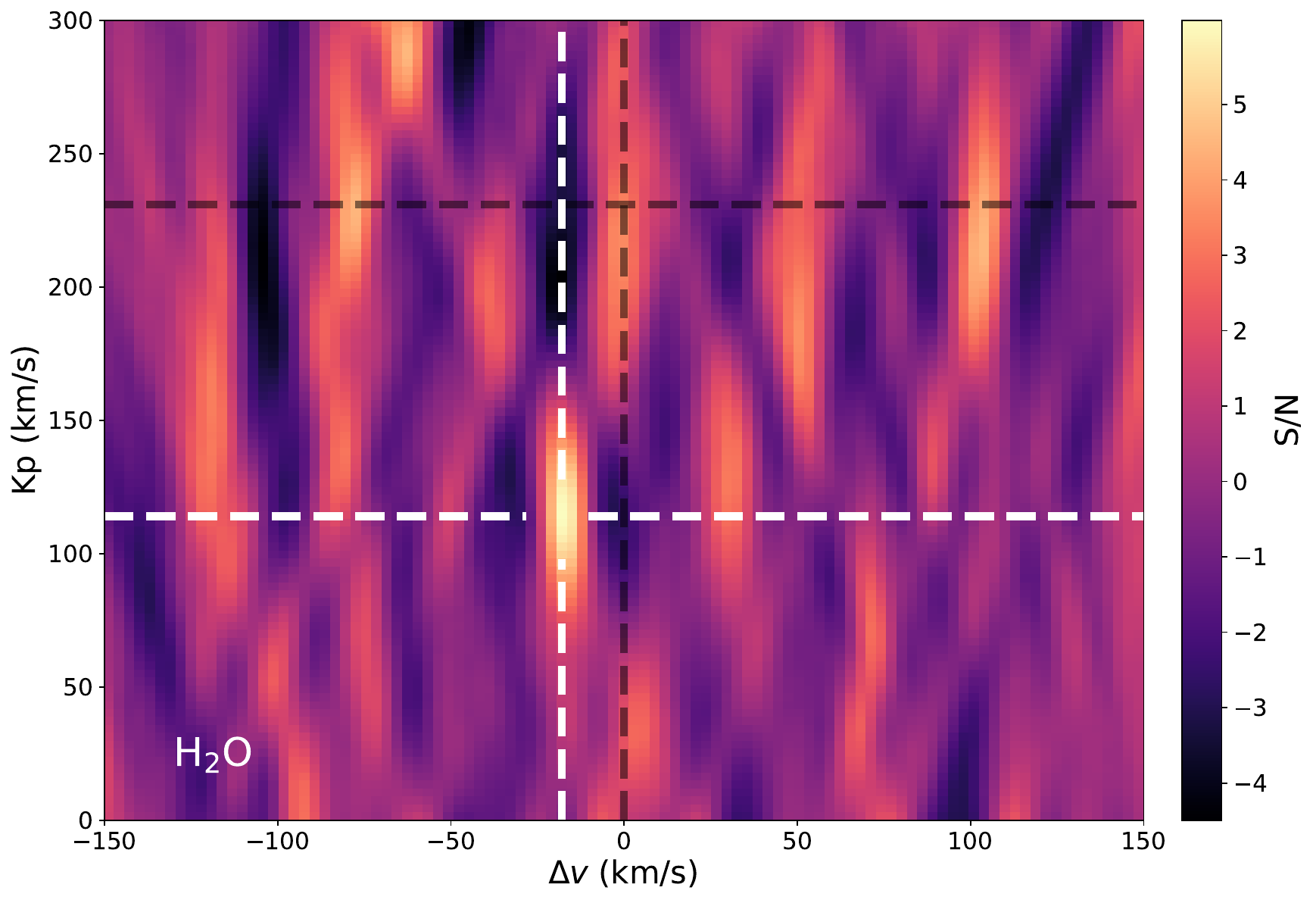}
\caption{$K_\mathrm{p}-\Delta \varv$ maps of H$_2$O for GIANO-B data ($\lesssim$1.75 microns) with two nights combined. The crossed white dashed lines indicate the S/N peak. The crossed black dashed lines indicate the expected $K_\mathrm{p}$ and $\Delta \varv$.}
\label{Fig_truncating_giano}
\end{figure}

We truncated the GIANO-B data at $\lesssim$1.75 microns (order 37 for GIANO-B) to repeat the CCF calculations. As shown in Figure~\ref{Fig_truncating_giano}, the result shows a lack of prominent signals at the expected locations when the wavelength coverage is insufficient.

\section{Simulation of High-resolution Phase-resolved Transmission spectra with nightside thermal Emission}\label{HRPCS}

\begin{figure}
\centering
\includegraphics[width=0.95\hsize,trim=20 0 0 40]{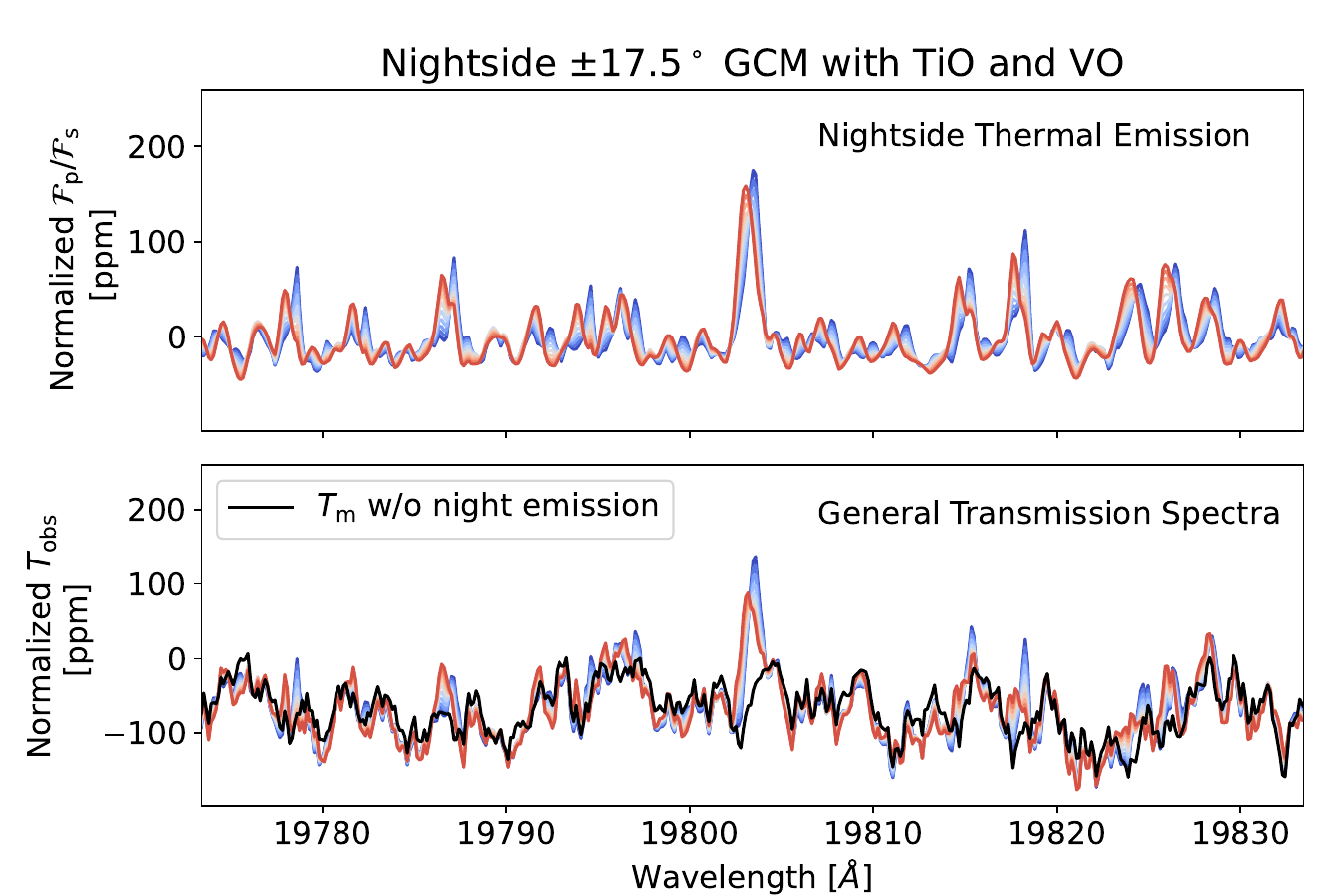}
\caption{An example of phase-resolved general H$_2$O transmission spectra simulation using a reference GCM with TiO and VO (i.e., nightside cooling is effective). \emph{Top panel}: The normalized phase-resolved nightside thermal emission spectra from $-$17.5$^\circ$ to +17.5$^\circ$ with a step of 3.5$^\circ$ (color-coded from blue to red).
\emph{Bottom panel}: Phase-resolved general transmission spectra with nightside thermal emission. The black line shows the classical transmission spectrum for H$_2$O lines (i.e., without considering night emission).}
\label{HRPCS_Figure}
\end{figure}

We conducted a phase-resolved high-resolution simulation of H$_2$O thermal emission spectra by post-processing the GCM outputs with PICASO \citep{2019ApJ...878...70B,2023ApJ...942...71M}. We initially tested a reference GCM from \citet{2024MNRAS.528.1016T}, including drag and H$_2$-H dissociation and recombination, with an equilibrium temperature of 2600 K, close to that of WASP-33b, and a host star temperature of 6500 K (lower than WASP-33). The reference models include TiO and VO, indicating that the nightside upper-layer cooling is effective. The other parameter settings are described in \citet{2024MNRAS.528.1016T}. We implemented radiative transfer calculations with 3D velocity fields by adapting the PICASO code with the radiative transfer methodology of \citet{1989JGR....9416287T}. For chemistry, we assumed chemical equilibrium without clouds (noting that the nightside of UHJs is typically covered with various high-temperature condensation clouds, to be discussed in a future paper). The line-by-line opacity database construction method can be found in the PICASO document\footnote{\url{https://natashabatalha.github.io/picaso/notebooks/10_CreatingOpacityDb.html}}. The rotation angle during the transit of WASP-33b is about 35.06$^\circ$ (i.e., the transit duration over the period). We calculated the phase-resolved thermal emission spectra within $\pm$17.5$^\circ$ of the mid-transit phase with a step of 3.5$^\circ$. The classical transmission spectrum was simply calculated by petitRADTRANS. We compared the strength of the classical transmission spectrum with the nightside emission spectra in Figure~\ref{HRPCS_Figure}.

\end{document}